\documentstyle[12pt]{article}

\newcommand{\jpsi}{$J/\psi$}
\newcommand{\xf}{$x_{\scriptscriptstyle F}$}
\newcommand{\alphas}{$\alpha_{\scriptscriptstyle S}$}
\newcommand{\et}{$E_{\scriptscriptstyle T}$}
\begin{document}
\begin{titlepage}
\begin{flushright}\vbox{\begin{tabular}{r}
           TIFR/TH/97-07\\
           March, 1997\\
           hep-ph/9703300\\
\end{tabular}}\end{flushright}
\begin{center}
   {\large \bf
      Is there Anomalous {\jpsi} Suppression\\
      in Present-day Heavy-ion Collisions?}
\end{center}
\bigskip
\begin{center}
   {R.\ V.\ Gavai\footnote{E-mail: gavai@theory.tifr.res.in} and
    Sourendu Gupta\footnote{E-mail: sgupta@theory.tifr.res.in},\\
    Theory Group, Tata Institute of Fundamental Research,\\
    Homi Bhabha Road, Bombay 400005, India.}
\end{center}
\bigskip
\begin{abstract}
   We re-analyse the data on {\jpsi} cross sections for proton-nucleus
   and nucleus-nucleus collisions obtained by the CERN experiments NA38,
   NA51 and NA50. Our systematic analysis of error propagation shows that
   the no anomalous suppression of {\jpsi} in Pb-Pb collisions can be
   substantiated at 95\% confidence limit. The significance of the NA50
   result would be enhanced by more precise p-A data on {\jpsi} cross
   sections.
\end{abstract}
\end{titlepage}

Recently the NA50 experiment at CERN announced an anomalous
suppression of {\jpsi} in Pb-Pb collisions at a beam energy of 158
GeV/nucleon \cite{na50,na50l}. Such an observation had earlier been proposed
as a signal for the formation of a quark-gluon plasma (QGP) \cite{helmut}.
In view of the implications of this observation, we reanalyse the CERN data
in this paper, taking into account many obvious sources of uncertainties. We
conclude that the data do not yet bear the claim they are asked to support.

The NA50 experiment as well as the NA51 and NA38 experiments have taken data
on {\jpsi} production in proton-nucleus collisions with beam energies
of 450 and 200 GeV, and in nucleus-nucleus collisions at
beam energies of 200 and 158 GeV/nucleon. The CM energies in equivalent
proton-proton collisions are $\sqrt S=29.1$, $19.4$ and $17.2$ GeV.
The acceptance region, $\delta$, for the data analysed in \cite{na50,na50l} is
$0\le y\le1$ (where $y$ is the rapidity of the {\jpsi} in the CM frame
of colliding protons at the same $\sqrt S$) and $|\cos\theta^*|\le0.5$
(where $\theta^*$ is the angle of the decay muon in the Collins-Soper frame).

The analysis consists of the following steps---
\begin{enumerate}
\item
   All the data on the cross section per nucleon ($\sigma/AA'$) are
   rescaled to 200 GeV beam energy, using a fit to previous data on
   the energy dependence of this cross section. The 158 Gev data are
   rescaled by a factor $1.32$ and the 450 GeV data by the factor $0.48$.
\item
   The rescaled data, at a common $\sqrt S=19.4$ GeV, are extrapolated
   to $|\cos\theta^*|\le1$, through a multiplicative factor of 2.
\item
   The resulting cross section is extrapolated to the full forward
   hemisphere, $y>0$, or equivalently, $x_{\scriptscriptstyle F}>0$
   by multiplying by $1.07$.
\item
   The rescaled $pA$ data in the forward hemisphere are fitted to a
   power-law
   \begin{equation}
      B\sigma({\cal A})\;=\;\sigma_0 {\cal A}^\alpha,
   \label{fit.form}\end{equation}
   where $B$ is the branching ratio for $J/\psi\to\mu^+\mu^-$, and
   ${\cal A}$ is the effective mass number, given by the product of
   the mass numbers of the projectile and target.
\end{enumerate}
It is found that the data obtained from light ions (O-Cu,
O-U and S-U) lie on the fitted line, whereas the Pb-Pb data lie far
below.  While the full procedure outlined above is used in \cite{na50l}, the
steps 2 and 3 are not performed in \cite{na50}. However, the extrapolation in 
$\sqrt S$ in both \cite{na50,na50l} uses a fit to the world data for 
$ x_{\scriptscriptstyle F} >0$ and full coverage in $\cos\theta^*$.

We note that there are uncertainties in every step of this procedure, and
also in the order in which these steps are to be performed. Each such
uncertainty is a source of error which must be combined with the statistical
and systematic errors in the Pb-Pb data. We present three seperate analyses.
First we accept the extrapolation in step 1 of the NA50 analysis \cite{na50}
and examine only the error in step 4 above. The Pb-Pb data is seen to 
be consistent with the p-A data at the 95\% confidence level (CL). Next 
we analyse the errors due to steps 1 and 4 only. Note that when the 
extrapolations are performed by NA50 \cite{na50}, the errors in the 
extrapolations in steps 2 and 3 are not needed in estimating the 
confidence limits of the final result.  Again the p-A and Pb-Pb data 
are consistent at the 95\% CL. Finally, we examine the situation when 
steps 2 and 3 are performed before step 1, and combine the errors in 
each of these four steps. The p-A and Pb-Pb data are once again seen 
to be consistent at the 95\% CL.

Since we shall repeatedly have to propagate errors, we outline the usual
statistical procedure for this \cite{intro}. If the expectation values of a
set of variables, $p_i$, are known along with their full covariance matrix,
$c_{ij}$, then the errors in any function of these variables, $f(p_i)$ can
be easily written down. Recall that the diagonal elements of the covariance
matrix are simply related to the errors, $\Delta p_i=\sqrt{c_{ii}}$. The
off-diagonal elements can be used to extract the normalised correlation
coefficients $r_{ij}=-c_{ij}/\sqrt{c_{ii}c_{jj}}$. These lie in the range
$|r_{ij}|\le1$. If $f(p_i)$ is a linear function of the $p_i$, then $\langle
f(p_i)\rangle= f(\langle p_i\rangle)$, and the error
\begin{equation}
   (\Delta f)^2\;=\;\sum_{ij} c_{ij}{\partial f\over\partial p_i}
                                    {\partial f\over\partial p_j}
\label{stat}\end{equation}
Note that when the correlations vanish, this reduces to the usual
formula for adding errors in quadrature\footnote{When $f$ is not a linear
function, the expectations and errors can be obtained either under the
assumption of a multivariate Gaussian distribution of the $p_i$, or by a
detailed bootstrap or Monte Carlo \cite{efron}. In this paper we will not
need to use a non-linear function $f$.}. In the succeeding analysis it
will turn out that we often have information on the full covariance matrix.
As a result, we can easily evaluate the formula in eq.\ (\ref{stat}). In
the exceptional cases when we have no knowledge of the off-diagonal elements,
we will choose them (by choosing $r_{ij}$) to minimise the errors. This
would give the best possible statistical significance to the result.

\begin{figure}
\vskip7truecm
\includegraphics{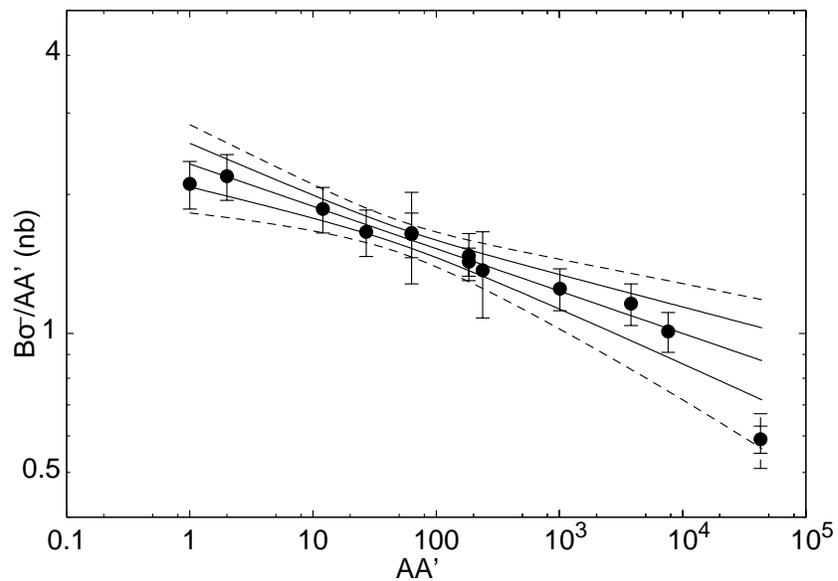}
\caption[dummy]{Data on the {\jpsi} cross section for $0<y<1$ and
   $|\cos\theta^*|\le0.5$, plotted against the product of the mass numbers
   of the target and projectile. The best fit power-law for the nuclear
   effects is shown along with the $1\sigma$ (enclosed by full lines),
   and the $2\sigma$ (dashed line).
   The error bar on the Pb-Pb point shows the 1 and $2\sigma$ errors
   with the same line styles.}
\label{fg.extrap.na50}\end{figure}

We begin by examining the errors involved in step 4 above. The form in
eq.\ (\ref{fit.form}) is linearised by taking the logarithm of both sides.
A fit is performed using the p-A data listed in \cite{na50l} obtained by
rescaling to a common $\sqrt S=19.4$ GeV. The result of this fit is---
\begin{equation}
   \sigma_0=2.33(1\pm0.11) {\rm\ nb},\quad
   \alpha=0.908\pm0.026,\quad
   {\rm Cov}(\log\sigma_0,\alpha)=-0.0026.
\label{fit.vals}\end{equation}
Note that the value of $\alpha$ is exactly the same as quoted
\cite{na50,na50l} by NA50.  Following exactly the NA50 analysis \cite{na50},
in Figure \ref{fg.extrap.na50} we show the best fit line to the $pA$ 
data along with the error bands about this fit line (using eq.\ \ref{stat}). 
The prediction for Pb-Pb collisions is $B\sigma(Pb{\rm -}Pb)=0.87(1\pm0.18)$
nb. The measured point, $B\sigma(Pb{\rm -}Pb)=0.59\pm0.04$ nb, is then
within $2\sigma$ of the extrapolation. Note that the extrapolation errors
are more crucial than the statistical errors on the Pb-Pb cross section.
As a result, the {\jpsi} suppression observed by
NA50 is indicative, but not a quantitative proof of the existence
of new physics, such as the QGP.

The assumption that O-Cu, O-U and S-U collisions do not yield a QGP can
be used to make a similiar statement. With this assumption, these three
extra points are included in the fit. Then we find
\begin{equation}
   \sigma_0=2.31(1\pm0.08) {\rm\ nb},\quad
   \alpha=0.911\pm0.014,\quad
   {\rm Cov}(\log\sigma_0,\alpha)=-0.001.
\label{fit.valsp}\end{equation}
The prediction for Pb-Pb collisions is $B\sigma(Pb{\rm -}Pb)=0.89(1\pm0.09)$
nb. The measured point is then less than
$3\sigma$ away from the prediction. The extra assumption makes the observation
stronger.

We turn now to the second analysis. Recall that the
rescaling of energy is performed using a fit to data on the $\sqrt S$
dependence of cross sections in the full forward hemisphere, using a formula
discussed later. We have no reason to believe that the two ratios
\begin{equation}\begin{array}{rl}
   R(\sqrt S,\sqrt S')\;=\;& \displaystyle
      {B\sigma(y>0,|\cos\theta^*|\le1,\sqrt S)\over
       B\sigma(y>0,|\cos\theta^*|\le1,\sqrt S')}\\
   r(\sqrt S,\sqrt S')\;=\;& \displaystyle
      {B\sigma(0<y<1,|\cos\theta^*|\le1,\sqrt S)\over
       B\sigma(0<y<1,|\cos\theta^*|\le1,\sqrt S')}
\end{array}\label{ratios}\end{equation}
are equal. The higher the energy, the larger is the allowed range
of $y$, and the slice $0<y<1$ is consequently a smaller part of the
full phase space. For the same reason, the orders in which the two
extrapolations are performed may give different results. However, cuts
on $\cos\theta^*$ are not expected to change these ratios provided they
are identical for the numerator and denominator.

\begin{figure}
\vskip7truecm
\includegraphics{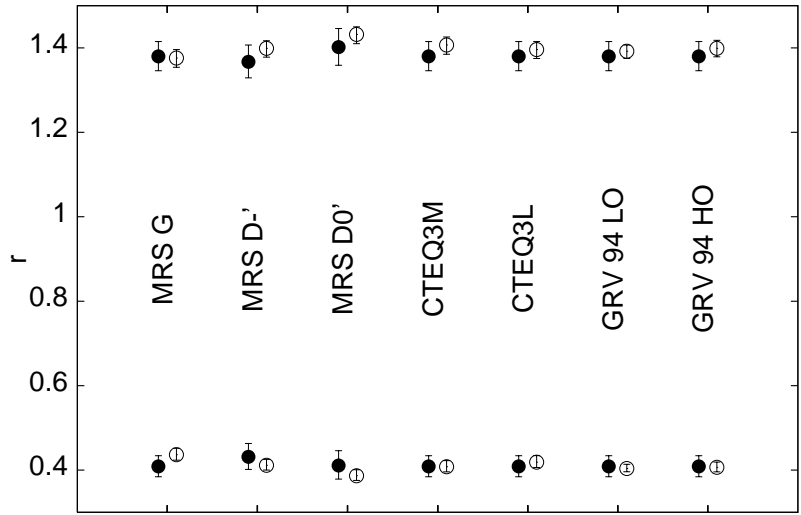}
\caption[dummy]{The scaling factor $r$ (eq.\ \ref{ratios}) in the colour
   octet model (full circles) and the colour evaporation model (open circles)
   for various structure functions. The upper set is for $r(19.4,17.2)$, and
   the lower for $r(19.4,29.1)$. The error bars show the variation due to
   varying all model parameters within one standard deviation and the charm
   quark mass between $1.4$ GeV and $1.6$ GeV.}
\label{fg.extrap.s}\end{figure}

Since we do not have data on the $\sqrt S$ dependence of $B\sigma_\delta$,
we use models to extract $r$. We have at our disposal the colour
evaporation model \cite{duality} and the colour octet model \cite{octet}.
Both give a reasonably good description of the $\sqrt S$ dependence of
$B\sigma_{y>0}$ as well as the distribution in $y$ (or {\xf}). For details
we refer the readers to
\cite{old1,old2}. In Figure \ref{fg.extrap.s} we show the ratio $r$
obtained in these models using several different sets of modern structure
functions. Note the good agreement among the models and the structure function
sets. The results shown in the figure translate to
\begin{equation}
   r(19.4,17.2)\;=\; 1.382_{-0.053}^{+0.064} \qquad{\rm and}\qquad
   r(19.4,29.1)\;=\; 0.412_{-0.034}^{+0.051},
\label{values}\end{equation}
where the central values are the averages over the model predictions and
the errors are the $1\sigma$ range of errors under variation of the structure
function set. These ratios and their errors may also be obtained from data
if they are available.

If the procedure followed by the NA50 group is adopted, then the errors
from the recaling of energy must be compounded with the statistical errors
on the data. We can add these errors in quadrature since the errors in $r$
are independent of the statistical errors in the NA50 and NA38 experiments.
In addition, the errors in the fits to the nuclear effects must be included
as before. The extrapolation to the full forward phase space is a common
multiplicative factor for all the data, and is unnecessary for the comparison.
Again our conclusions will therefore apply to both Refs. \cite{na50} and
\cite{na50l}.  Using this procedure, we find that the extrapolation of 
the p-A data to Pb-Pb is $1.0(1\pm0.21)$ nb, whereas the measured value 
after rescalings is $0.61\pm0.04$ nb. Thus the Pb-Pb data is within 
$2\sigma$ of the extrapolation from the p-A data.

We move on to the third analysis. Recall that
the Collins-Soper frame is the rest frame of the decaying particle with
the $z$-direction chosen such that the spin is aligned in this direction
if the helicity is positive. Then, in this frame, the angular distribution
of the decay leptons coming from any massive vector particle (either
$\gamma^*$ for the Drell-Yan process or any vector quarkonium) should be
$1+\cos^2\theta^*$, when strong interaction corrections are neglected. Then
the extrapolation from $|\cos\theta^*|\le0.5$ to the full acceptance is
a factor $32/13$ for both Drell-Yan and vector quarkonium
production, independent of $\sqrt S$.

The NA50 group uses this factor for Drell-Yan cross sections, but a factor
2 for {\jpsi}.
Since this is an overall factor, it does not affect the present analysis.
However, in any comparison of Drell-Yan and {\jpsi} cross sections, the
correct factor\footnote{ After this paper was written we were informed by some
members of the NA50 collaboration that data on $J/\psi$ production
at lower energies shows a flat angular distribution, suggesting 2 as
the appropriate rescaling factor.}
must be used. It should also be recalled that QCD corrections
change this distribution. The modification can be written as a series in the
strong-coupling, {\alphas}, multiplying $\cos^2\theta^*$. The first term in
this series is 1 for all vector particles, but the subsequent terms may be
different for $\gamma^*$ and {\jpsi}. Then the extrapolation uncertainty in
the ratio of Drell-Yan and {\jpsi} cross sections is of the order of \alphas.

There remains the question of extrapolation from $0<y<1$ to $y>0$ (or
equivalently, $x_{\scriptscriptstyle F}>0$). Many experiments \cite {expts}
have parametrised the {\xf} distribution in the form
$(1-x_{\scriptscriptstyle F})^d$. Using this form, and the range of errors
quoted, we find that
\begin{equation}
   B\sigma(y>0) \;=\; B\sigma(0<y<1) \times
     \left\{
        \begin{array}{lr}
         1.03-1.05 & (\sqrt S=16.8)\\
         1.06-1.07 & (\sqrt S=20.4)\\
         1.11-1.26 & (\sqrt S=27.4)\\
        \end{array}
     \right.
\label{extrap.y.expt}\end{equation}
In order to obtain these scalings at the values of $\sqrt S$ we need, we could
use a parametrisation of the $\sqrt S$ dependence of the exponent $d$ 
fitted in the form \cite{schuler}
\begin{equation}
   d\;=\; {a\over1+b/\sqrt S}.
\label{extrap.xf}\end{equation}
However, this gives results at variance with those in
eq.\ (\ref{extrap.y.expt}).
This might be due to fitting too simple a functional form to the data.
On the other hand, the colour octet model calculations yield numbers in
reasonable agreement with eq.\ (\ref{extrap.y.expt}) and the actual value
used by the NA50 analysis \cite{na50l} at $\sqrt S=19.4$ GeV. We find
\begin{equation}
   B\sigma(y>0) \;=\; B\sigma(0<y<1) \times
     \left\{
        \begin{array}{lr}
         1.053\pm0.009 & (\sqrt S=17.2)\\
         1.082\pm0.014 & (\sqrt S=19.4)\\
         1.224\pm0.041 & (\sqrt S=29.1)\\
        \end{array}
     \right.
\label{extrap.y}\end{equation}

The third analysis may be performed by doing the extrapolation from $\delta$
to the forward hemisphere first, and then using the available data on the
energy dependence of $B\sigma_{y>0}$ to bring all the data to a common value
of $\sqrt S$. In this case the errors in eq.\ (\ref{extrap.y}) must be
combined with the error in energy extrapolation.

The rescaling of $B\sigma$ with $\sqrt S$ is performed by a fit of the form
\cite{schuler}
\begin{equation}
   B\sigma(\sqrt S)\;=\;\sigma_0\left(1-{M_{J/\psi}\over\sqrt S}\right)^n,
\label{extrap.e}\end{equation}
to data on the total forward {\jpsi} cross section at different $\sqrt S$.
The problem is linearised by taking logarithms on both sides of this
equation. Taking $M_{J/\psi}=3.097$ GeV, \cite{schuler} reports the best fit
to be\footnote{Using the data collected in Table 8 of \cite{schuler} and
using MINUIT without linearising the problem, we get somewhat different values
of the parameters--- $\sigma_0=44_{-11}^{+17}$ nb and $n=13.0_{-1.3}^{+1.9}$.
This evidently increases the errors quoted below on the quantity $R$.}
\begin{equation}
   \sigma_0\;=\;38(1\pm0.16) {\rm\ nb},\qquad{\rm and}\qquad
   n\;=\;12.0\pm0.9.
\label{extrap.e.fit}\end{equation}
These numbers were obtained by fitting both p-p and p-A data. Nuclear effects
in the p-A data were eliminated by assuming $\alpha=0.91$ in \cite{schuler}.
The formula in eq.\ (\ref{extrap.e}), with the values in eq.\ 
(\ref{extrap.e.fit}), was used by the NA50 group to compute the rescaling in
$\sqrt S$ \cite{na50,na50l}. We find
\begin{equation}
   R(19.4,17.2)\;=\;1.34\pm0.03 \qquad{\rm and}\qquad
   R(19.4,29.1)\;=\;0.48\pm0.03,
\label{extrap.e.vals}\end{equation}
where the error in $R$ depends only on the error in $n$.
Possible systematic errors in this extrapolation can be investigated by
changing the form in eq.\ (\ref{extrap.e}). In particular, at larger
energies, one should take into account the expected rise in cross sections
as (roughly) $\log S$.

\begin{figure}
\vskip7truecm
\includegraphics{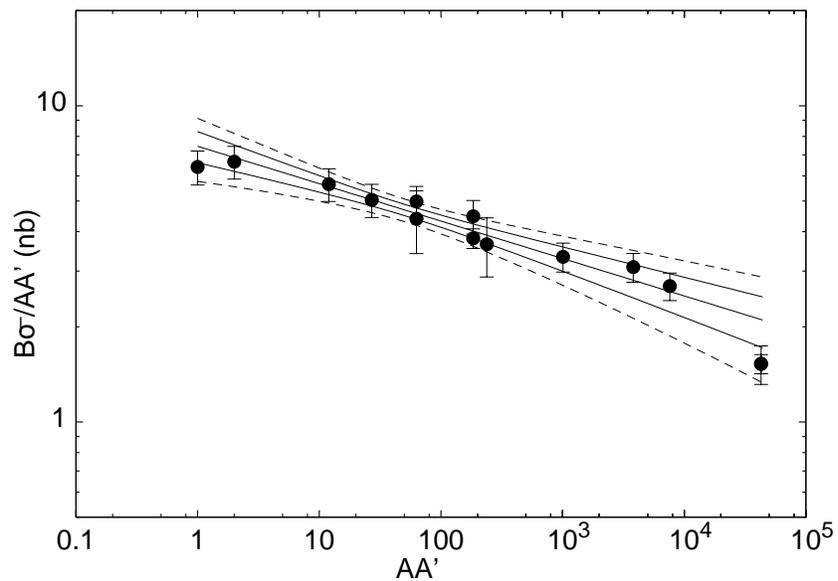}
\caption[dummy]{Data on the {\jpsi} cross section $B\sigma(y>0,19.4)$
   plotted against the product of the mass numbers of the target and
   projectile. The best fit power-law for the nuclear
   effects is shown along with the 1 and $2\sigma$ error bands (full and
   dashed lines respectively). Similiarly, the 1 and $2\sigma$ errors on
   the Pb-Pb point are also shown.}
\label{fg.extrap.ours}\end{figure}

Using eqs.\ (\ref{extrap.y}) and (\ref{extrap.e.vals}), along with a scaling
factor of $32/13$ for the extrapolation of the Collins-Soper angle, we find
that
\begin{equation}\begin{array}{rl}
  B\sigma(y>0,|\cos\theta^*|\le1,19.4) &\;=\; (1.41\pm0.02)\times B\sigma_\delta(17.2)\\
  B\sigma(y>0,|\cos\theta^*|\le1,19.4) &\;=\; (0.59\pm0.02)\times B\sigma_\delta(29.1).
\end{array}\label{factor}\end{equation}
Recall that the factor of $32/13$ is a common overall factor, and hence its
value or error is immaterial to the following analysis.
The errors in the coefficients in eq.\ (\ref{factor}) are independent of the
measurements of the NA38, NA50 and NA51 experiments. Hence they may be added
in quadrature to the statistical errors on the data. When the $pA$ data are
scaled in this fashion, the fit of the form in eq.\ (\ref{fit.form}) becomes
\begin{equation}
   \sigma_0=7.45(1\pm0.11) {\rm\ nb},\quad
   \alpha=0.882\pm0.026,\quad
   {\rm Cov}(\log\sigma_0,\alpha)=-0.0028.
\label{fit.valfin}\end{equation}
Although the value of $\alpha$ decreases, it still remains
consistent with the E-789 extraction of $\alpha=0.90\pm0.02$
\cite{e789}. The scaling of the CERN data, however, is a little stronger
than that observed by the E-772 experiment\footnote{The comparison of all
these values of $\alpha$ is a measure of the systematic uncertainty in
this quantity. One can assign the value $0.02$ to the systematic error on
$\alpha$.} which sees $\alpha=0.920\pm0.008$
\cite{e772}.
The comparison of the extrapolation of this data with the Pb-Pb cross
section is shown in Figure \ref{fg.extrap.ours}. Note that the cross
section for {\jpsi} production in Pb-Pb collisions is still consistent
with those for p-A collisions at the $2\sigma$ level.

If the light-ion data is included in the fit to the form in eq.\ 
(\ref{fit.form}), by making the assumption that there is no new physics
seen with light-ions, then the fitted parameters become
\begin{equation}
   \sigma_0=7.23(1\pm0.08) {\rm\ nb},\quad
   \alpha=0.890\pm0.014,\quad
   {\rm Cov}(\log\sigma_0,\alpha)=-0.0011.
\label{fit.valfin.aa}\end{equation}
With this extra assumption, the extrapolated line fails to agree with
the Pb-Pb data at the $2\sigma$ level, but remains consistent at the
$3\sigma$ level.

The experiments have more information than just the total cross section. In
particular, they measure the transverse energy, {\et}, dependence of the
{\jpsi} cross section.
A quantitative analysis \cite{satz} found that the well-known absorption
mechanism can account for the {\et} dependence in S-U collisions, but not
in {Pb-Pb} collisions. This analysis uses the survival probability of the
{\jpsi}, defined by the ratio
\begin{equation}
   S(E_{\scriptscriptstyle T})\;=\; \left({2\over AA'}\right)
    {\sigma^{AA'}(y>0,|\cos\theta^*|\le1,\sqrt S=19.4,E_{\scriptscriptstyle T})
             \over
    \sigma^{pD}(y>0,|\cos\theta^*|\le1,\sqrt S=19.4,E_{\scriptscriptstyle T})},
\label{surv}\end{equation}
where the cross sections have been scaled to a common $\sqrt S$ and the full
forward hemisphere, by the prescription of NA50 \cite{na50l}. It was found
that, for Pb-Pb collisions, the data point at the lowest {\et} agrees with
a Glauber description, but the remaining points fall below the prediction
\cite{satz}.

\begin{figure}
\vskip7truecm
\includegraphics{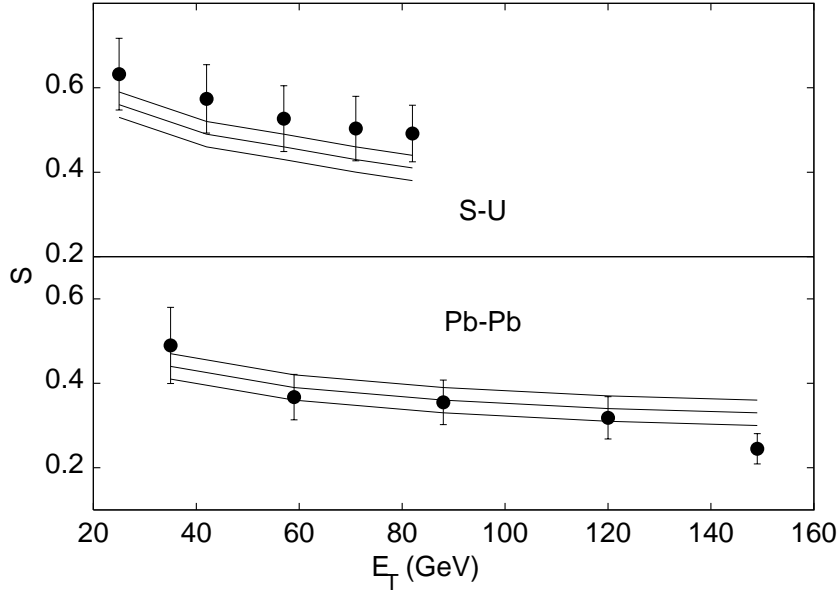}
\caption[dummy]{Data on the {\et} dependence of the {\jpsi}
   survival probability $S$ (eq.\ \ref{surv}). The band shows the
   prediction of a Glauber model \cite{satz} within a $1\sigma$
   variation of the model parameters.}
\label{fg.etdep}\end{figure}

This analysis could also be redone in a fashion similiar to our analysis
of the total forward cross sections. Figure \ref{fg.etdep} shows the data
points which are scaled as in the second analysis (eq.\ \ref{values}). All
data agree with the Glauber model at the $1\sigma$ level. The fit to the
S-U data is not as good as before; although remaining consistent at the
$1\sigma$ level it seems to lie systematically above the Glauber model
prediction. The Pb-Pb data agree much better, with the sole exception of
the point at highest {\et}. This point agrees with the model only at the
$2\sigma$ level. When eq.\ (\ref{factor}) is used for the rescalings,
both the S-U and Pb-Pb data lie systematically below the Glauber model
predictions, although being consistent at the $2\sigma$ level. In either
case, the Glauber model does equally well (or badly) in explaining the
{\et} dependence of both the S-U and Pb-Pb data.

In conclusion, we find little statistical significance for an anomalous
suppression of {\jpsi} cross section in Pb-Pb collisions at the CERN SPS
energy. In our analysis, the uncertainties are dominated by the errors
due to the extrapolation of the usual nuclear effects in p-A collisions
to Pb-Pb. Clearly, reduced errors on the p-A cross sections would enhance
the statistical significance of the deviations seen by NA50. The importance
of controlling nuclear effects was first pointed out in \cite{ours}; it
continues to be an important problem. Even in future
experiments at the RHIC and LHC, the p-A data will always serve as a baseline
of ``known physics''. We feel that the worldwide heavy-ion effort must be
supplemented by an effective study of the systematics of p-A collisions.
Unless improved data from p-A collisions are available, interpretations
of data from present and future heavy-ion colliders may be plagued by
uncertainties of the kind we have pointed out.

\end{document}